\documentclass[11pt]{article}
 
\usepackage{hyperref}
\usepackage{amsmath,amsfonts,amssymb}
\hypersetup{dvips,dvipdfm,colorlinks=true,urlcolor=magenta,filecolor=magenta,linktoc=page,citecolor=red,linkcolor=blue,bookmarks=true}
\usepackage{array}
\usepackage{graphicx,epstopdf}
\usepackage{multirow}
\usepackage{xcolor}
\begin{document}
	\title{Quasinormal modes and Grey body factors of Wormholes: From General prescription to Einstein-Gauss-Bonnet realizations}
		\author{Madhukrishna Chakraborty$^1$ \footnote{\url{ chakmadhu1997@gmail.com}} ~~and ~~Subenoy Chakraborty$^2$ \footnote{\url{schakraborty.math@gmail.com} (corresponding author)}}
	\date{%
		\scriptsize	$^1$Department of Mathematics, Techno India University, Kolkata-700091, West Bengal, India\\%
		$^2$Department of Mathematics, Brainware University, Kolkata-700125, West Bengal, India\\
		$^2$ Shinawatra University, 99 Moo 10, Bangtoey, Samkhok, Pathum Thani 12160, Thailand\\
		$^2$ INTI International University, Persiaran Perdana BBN, Putra Nilai, 71800 Nilai, Malaysia\\[2ex]%
		{}
	}
	\maketitle
	\begin{abstract}
		 Traversable wormholes (TWHs) are one of the most exciting predictions of General Relativity (GR) that offer short-cuts through space-time. However, their feasibility requires the violation of the null energy condition (NEC) and make their detection a bit difficult. The paper aims to show the new avenues delving deep into the observational prospects of TWHs via quasinormal modes (QNMs) and grey body factors (GBFs). These are the two elementary aspects of wave dynamics. Given their distinct spectral imprints, these features provide a potential means to distinguish wormholes from black holes in gravitational wave observations. The role of QNMs in characterizing the ringdown phase of perturbations and the GBFs in determining transmission probabilities through wormhole barriers have been explored by a general description and then fed to Einstein-Gauss-Bonnet (EGB) WH solutions in isotropic as well as anisotropic cases. Finally, a correspondence of the QNM frequencies with radius of the WH shadows have been made and effect of Gauss Bonnet parameter in the QNM spectra has been discussed.
	\end{abstract}
	
	\section{Introduction}
Wormhole (WH) is a fascinating yet a hypothetical astrophysical object, found as a solution of Einstein's field equations and has been considered topologically as a shortcut passage between two distant universes or two {\color{red} distant} space-time regions of the same Universe \cite{flamm}. The existence of a TWH requires matter that violates the NEC, particularly in the vicinity of the throat \cite{Einstein:1935tc}, \cite{ellis}. Immediately, after the formulation of Einstein gravity Flamm \cite{flamm} gave a solution to the field equations-the first ever known theoretical prediction of WH. Later, Einstein and Rosen \cite{Einstein:1935tc} gave another solution having bridge like nature and is commonly known as Einstein-Rosen bridge. The idea of traversability introduced by Morris-Thorne is very popular due to simplified and physically transparent nature \cite{Morris:1988cz}. They argued that traversability can be ensured by exotic matter surrounding the throat of the WH. As a time-like or null test particle can pass through the TWH in finite time so one may observe their response in the form of the wave scattering and quasinormal ringing or what we say as ringdown.

The static spherically symmetric Lorentzian traversable wormhole (LTWH) introduced by Morris-Thorne is characterized by its shape function $b(r)$ and red-shift function $\Phi(r)$. An extension to axially symmetric TWH was introduced by Teo \cite{Teo:1998dp}. Being an astrophysical entity, the interesting feature of WH is the light deflection character both in strong as well as in the weak deflection limit for a WH space-time \cite{Perlick:2015vta}. Interestingly, WHs behave like BHs in the context of photon trajectories around them. It is well known that around the BHs, the photons can either fall into the BH or scattered away from it to the infinity. In between there are critical photon orbits which separate the above two sets and are termed as unstable photon orbits. They appear to a distant observer as shadow \cite{Nedkova:2013msa}. The situation should be similar in case of WHs and one should have WH shadow. In the context of BH, there is a strong observational evidence regarding shadows. This is because, the shadows of $M87*$ and $SgrA*$ have been found by Event Horizon Telescope (EHT) \cite{Afrin:2022ztr}. As a result, the astrophysicists are very excited to have an observational evidence for other compact objects including WH. From perturbative point of view, the WHs are gravitationally stable, provided TWH is supported by some exotic matter (dark energy). It is speculated that in course of inflationary scenario the primordial microscopic WHs evolve to macroscopic size. Also, assuming existence of WHs one may observe new stars considering WH as a bubble. In analogy to supermassive BHs, accretion disk should be visible across a rotating WH surrounded by some luminous matter. Hence, the detection of WHs may be possible through indirect evidences like gravitational lensing, unique signatures in quasinormal modes and/ or anomalous energy distributions inconsistent with Black Hole (BH) models. The century waited theoretical prediction of GR has been observationally verified by LIGO and VIRGO collaborators \cite{LIGOScientific:2017vwq}, \cite{LIGOScientific:2016aoc}. It is well known today that in GWs produced during the collision of BHs, the final stage of the ringdown phase is characterized by the quasinormal modes \cite{Patra:2024srh}. The study of these QNMs reveal that it is possible to have such QNMs due to the existence of WHs also \cite{Konoplya:2025mvj}. The results of the detection of GW not only prove the existence of BH but also hints a distinction between a BH and WH or a possible collision between a BH and a WH \cite{Konoplya:2025mvj}. Interestingly, since the discovery of GW, astrophysicists have been trying to associate the observational aspects with the theoretical predictions of compact objects. Specifically, the real part of the QNMs are related to the angular velocity of the last circular null geodesics and a correspondence between the QNMs and the strong lensing limit. Subsequently, it has been shown that the shadow radius is related to the real part of QNMs not only for static/rotating BHs but also is true for static/ rotating WHs --- a nice interplay between GW astronomy and the shadow of compact objects. More specifically, for a distant observer the GW from a compact object behaves as massless scalar field propagating along the last unstable null orbit and gradually approaching to spatial infinity.

In the recent years there have seen significant progress in distinguishing WHs from BHs through wave dynamics and observational probes. In particular, the study of QNMs and greybody factors in alternative theories of gravity has been widely pursued after the first detection of GWs. Konoplya, Zhidenko and collaborators have investigated how deviations in QNM spectra may serve as evidence of non-black-hole compact objects, highlighting possible late-time echoes in gravitational wave signals as observational discriminants. Such echoes and ringdown deviations are now actively explored as potential evidence for WH-like geometries in the strong gravity regime \cite{Patra:2024srh}, \cite{Konoplya:2025mvj}, \cite{Konoplya:2024vuj}, \cite{Konoplya:2024lir}, \cite{Konoplya:2023moy}, \cite{Fernandes:2020rpa}. Moreover, recent works have emphasized stability analyses of WH solutions in higher-curvature theories such as Einstein–Gauss–Bonnet (EGB), Lovelock, and beyond, showing that stability constraints strongly restrict the parameter space where QNMs yield physically meaningful spectra       \cite{Fernandes:2020rpa}, \cite{Kumar:2018ple}. On the observational front, Afrin, Vagnozzi and Ghosh (2023) have demonstrated how shadow size and QNM frequencies can be correlated in order to test WH metrics against Event Horizon Telescope (EHT) data \cite{Afrin:2022ztr}. In parallel, studies by Bugaev et al. (2021) \cite{Bugaev:2021dna}, Neto et al. (2023) \cite{Neto:2022pmu}, and Bolokhov and Skvortsova (2024) \cite{Bolokhov:2024otn} have deepened the correspondence between WH shadows, quasinormal ringing, and transmission probabilities, making greybody factors a promising tool for distinguishing astrophysical WHs from classical black holes \cite{Mehdizadeh:2021kgv}. Alongside this, Maldonado-Villamizar et al. (2022) and Zanoletti et al. (2024) \cite{Zanoletti:2023ori}-\cite{Feng:2020duo} have examined perturbations in EGB backgrounds, providing explicit links between greybody spectra and modified dispersion relations. Finally, multi-messenger constraints are increasingly important. LIGO–VIRGO detections have been used to constrain exotic compact objects, and several analyses  like Konoplya et al. \cite{Konoplya:2024lir}; Mehdizadeh and Ziaie \cite{Mehdizadeh:2021kgv} argue that systematic deviations in the ringdown phase may be used to discriminate wormholes from black holes in upcoming high-sensitivity gravitational-wave observations. These developments underscore that observational prospects for wormholes are not purely theoretical, but increasingly testable. By situating the present study of QNMs and GBFs in EGB WHs within this context, we highlight its relevance to ongoing efforts in GW astronomy and black-hole mimicker phenomenology.

	The motivation to choose Einstein- Gauss- Bonnet (EGB) theory \cite{Zanoletti:2023ori}-\cite{Feng:2020duo} is due to its high curvature nature and also its relation to Lovelock gravity \cite{Dehghani:2009zza}-\cite{Lovelock:1971yv}. It is well known that EGB theory is a particular case of Lovelock gravity which is a natural generalization of Einstein gravity to higher dimensions. This higher dimensional Lovelock gravity theory was introduced long back by Lanezos \cite{Lanczos:1938sf} and then later it was again introduced by Lovelock \cite{Lovelock:1971yv}. On the other hand, Einstein Gauss Bonnet (EGB) modified theory of gravity has been used extensively for (i) its association with low energy limit of string theory \cite{Zwiebach:1985uq}, (ii) absence of instability when expanded about flat space-time \cite{Wheeler:1985nh}, (iii) its association with ghost-free nontrivial gravitational self-interactions \cite{Nojiri:2018ouv}.  It was well known that EGB theory is valid for $D>4$. If $D=4$, the EGB theory becomes topological due to the total derivative form of the Lagrangian and hence it poses no effect in gravitational dynamics. This limitation of EGB theory was addressed recently by Glavan and Lin \cite{Glavan:2019inb}, where a rescaling of the coupling parameter $\alpha$ has been taken into account with the limit as $D\rightarrow4$. Consequently, one has a non-trivial dynamics in 4D EGB gravity theory.  However, it is to be noted that the original regularisation proposed by Glavan and Lin \cite{Glavan:2019inb} has provoked an active debate regarding its foundational status and whether it defines a genuinely four-dimensional gravitational theory. Several works have argued that the naive $D\!\to\!4$ limiting procedure is not well-posed and that the resulting field equations cannot be obtained from a covariant four-dimensional action in the usual metric formalism (see e.g. Gürses, Şişman \& Tekin \cite{Gurses:2020ofy}; Ai, Hennigar et al. \cite{Hennigar:2020lsl};  Fernandes et al. \cite{Fernandes:2020rpa}). At the same time, alternative regularisation/embedding approaches and scalar–tensor reformulations have been proposed which lead to different (but related) four-dimensional effective theories that avoid some of these criticisms (see e.g. Aoki, Gorji \& Mukohyama \cite{Aoki:2020lig}; and the scalar–tensor reinterpretations discussed in the literature). For this reason, we employ the regularized 4D EGB background here strictly as a phenomenological framework to explore the qualitative imprint of the Gauss–Bonnet coupling on wormhole shadows, quasinormal modes and grey-body factors; a full assessment of formal consistency and the relation to alternative, explicitly covariant 4D reductions is beyond the scope of this work and left to future study. There are a couple of interesting features of this 4D gravity theory namely, (i) it is not linked with the conclusions of Lovelock's theorem and (ii) it is free of Ostrogradsky instability.  Moreover, the static spherically symmetric vacuum black-hole solutions in 4D EGB gravity is interesting in the sense that (a) the black-holes are free from singularity at small distances, (b) due to repulsive nature of gravity at small distances, an infalling  particle fails to encounter the singularity \cite{Kumar:2020xvu}.

The layout of the paper is as follows: Section 2 gives a general prescription of WH shadows, QNMs and GBFs. Section 3 deals with the analysis of QNMs and GBFs in EGB WHs. Section 4 gives a semi analytic approach to justify the QNM-shadow correspondence presented in the paper. Finally, the paper ends with conclusion in section 5.
\section{Shadows, QNMs and GBFs of TWHs: a general prescription}
\subsection{TWHs and their shadows}
A general spherically symmetric static WH space-time of the Morris Thorne class (traversable) is given by the 	line element \cite{Ghosh:2021msy}-\cite{Halder:2019urh}
\begin{equation}
	ds^{2}=-e^{2\Phi(r)}dt^{2}+\dfrac{1}{\left(1-\dfrac{b(r)}{r}\right)}dr^{2}+r^{2}d\Omega_{2}^{2}\label{eq9*}
\end{equation}
where $d\Omega_{2}^{2}=d\theta^{2}+\sin^{2}\theta d\phi^{2}$, $b(r)$ is called the shape function that determines the shape of the WH. $\Phi(r)$ is called the red-shift function and it is a function of the radial coordinate $r$ such that $r_{0}\leq r<\infty$. Red-shift function is always finite everywhere to avoid horizon or singularity and it is used to detect the red-shift of the signal by a distant observer giving information about the radial tidal force. $r_{0}$ is the radius of the WH throat. There are certain restrictions on $b(r)$. They are as follows:
\begin{enumerate}
	\item $b(r_{0})=r_{0}$ is the throat condition and $b(r)<r$ for $r>r_{0}$ (metric condition).
		\item Asymptotic flatness: $\dfrac{b(r)}{r}\rightarrow 0$ as $r\rightarrow \infty$.
	\item Flaring out condition: $b(r)-rb'(r)>0$ where $b'(r)=\dfrac{db(r)}{dr}$.
\end{enumerate} 
The ECs that play a crucial role in the formation of a feasible WH configuration are given by \cite{mc}
\begin{enumerate}
		\item Strong Energy Condition (SEC):  $\rho+p_{r}\geq0$, $\rho+p_{t}\geq0$, $\rho+p_{r}+2p_{t}\geq0$.
			\item Weak Energy Condition (WEC): $\rho\geq0$,  $\rho+p_{r}\geq0$, $\rho+p_{t}\geq0$.
	\item Null Energy Condition (NEC): $\rho+p_{r}\geq0$, $\rho+p_{t}\geq0$.
	\item Dominant Energy Condition (DEC): $\rho-|p_{r}|\geq0$, $\rho-|p_{t}|\geq0$.
\end{enumerate}

From a mathematical perspective, the presence of Lorentzian TWHs might be inferred through the behavior of test fields near the WH, specifically via wave scattering and quasinormal mode oscillations. Advanced observational methods—such as high-resolution interferometry, precise timing of astrophysical phenomena, and multi-messenger astronomy—could potentially detect these effects. However, the necessity of exotic matter and the strict limitations imposed by stability considerations pose significant challenges to their observational verification. The discovery of a WH would have profound implications, potentially reshaping and broadening our understanding of spacetime topology, cosmic connectivity thereby offering insights into quantum gravity and the deeper structure of the Universe. This would possibly even enable our understanding regarding the concepts of time travel, interstellar and intergalactic transport.

Suppose among the distant space-time regions that are connected by a TWH, one space-time region is illuminated by a light source while there is no light source near the throat  of the WH in the space-time region on the other side. Usually, the photons have two possible fates \cite{Bugaev:2021dna}
(a) photons enter the WH and traverse along the throat, 
(b) photons scatter away from the WH to infinity.

As a result, a distant observer located in the first region will only detect photons that are scattered, while those that are captured by the wormhole will be absent from view, creating a dark region against the bright background. This dark region is known as the shadow of the wormhole. From a mathematical standpoint, for photons to be scattered, their radial trajectory must have a turning point defined by $\dfrac{dr}{d\lambda}=0$ i.e, $V_{eff}=0$. The critical orbit that separates scattered paths from those that plunge into the wormhole corresponds to the maximum of the effective potential. This orbit is spherical and inherently unstable—any small disturbance can cause the photon to either escape or be captured. Thus, the critical orbit is characterized by $V_{eff}=0=\dfrac{dV_{eff}}{dr}$ and $\dfrac{d^{2}V_{eff}}{dr^{2}}\leq 0$. So for WH configuration, the radius of the photon sphere $r_{ph}$ which locates the apparent image of the photon rings satisfies \cite{Bugaev:2021dna}-\cite{Neto:2022pmu}
\begin{equation}
	r\Phi'(r)=1
\end{equation} i.e, 
\begin{equation}
	e^{\Phi(r)}=\Phi_{0} r
\end{equation}
In fact, $r_{ph}$ is the largest real root of the equation $r\phi'(r)=1$ and it identifies the shadow radius as
\begin{equation}
	r_{sh}=r e^{-\Phi(r)}|_{r=r_{ph}}\label{eq75}
\end{equation} where $r_{sh}$ is the radius of the shadow. For convenience, we assume that the observer is situated far away from the WH so that the radius of WH shadow can be expressed as \cite{Bugaev:2021dna}
\begin{equation}
	r_{sh}=r_{ph}e^{-\Phi(r_{ph})}=\sqrt{X^{2}+Y^{2}}
\end{equation} where in the observer's frame $(X,Y)$ are identified as the celestial coordinates. Now, let $R_{0}$ be the distance of the observer from the WH and $\theta_{0}$ be the observer's angular coordinate (or inclination angle) i.e, $(R_{0},\theta_{0})$ is the location of the observer, then $(X,Y)$ essentially identifies the boundary curves of the WH shadow (i.e, the apparent shape of the shadow) and are related to the observer's coordinates $(R_{0},\theta_{0})$ as \cite{Bugaev:2021dna}-\cite{Neto:2022pmu}
\begin{eqnarray}
	X= \lim_{R_{0}\rightarrow \infty}(-R_{0}^{2}\sin \theta_{0})\dfrac{d\phi}{dr}\\
	Y= \lim_{R_{0}\rightarrow \infty}\left(R_{0}^{2}\dfrac{d\theta}{dr}\right)
\end{eqnarray}
i.e,
\begin{eqnarray}
	X=-\dfrac{\mu}{\sin \theta_{0}}\\
	Y=\sqrt{\nu-\mu^{2}/\sin^{2}\theta_{0}}
\end{eqnarray}
where (expressions of the four velocities have been used) $\lim_{r_{0}\rightarrow \infty}e^{-2\Phi(R_{0})}=$ constant is considered and $\mu,~\nu$ are called impact parameters \cite{Bugaev:2021dna}-\cite{Neto:2022pmu}. Using the above interrelations among the celestial coordinates and the impact parameters one may form the shadow of the TWH. For a diagrammatic representation of the shadow one has to plot $X$ vs $Y$ to identify the shadow's boundary (for simplicity one may choose the equatorial plane $\theta_{0}=\dfrac{\pi}{2}$).

\subsection{QNMs and GBFs of TWHs}
 Quasinormal modes (QNMs) and greybody factors are fundamental in understanding wave dynamics in the arena of curved spacetime, particularly near compact astrophysical objects like black holes (BHs) or wormholes (WHs). QNMs represent the intrinsic oscillatory responses of such objects to external perturbations such as gravitational waves (GWs). While, greybody factors quantify the partial transmission and reflection of radiation due to spacetime curvature. For spherically symmetric, asymptotically flat or de Sitter BHs, a correspondence between QNMs and greybody factors has been well established using the WKB approximation \cite{Rosato:2025rgy}. Interestingly, greybody factors tend to remain more stable under near-horizon geometric deformations, whereas QNM overtones are highly sensitive to such changes \cite{gb}. Given that the boundary conditions for QNMs and greybody factors are similar in both BH and WH spacetimes, it is reasonable to extend this correspondence to WHs as well. The correspondence between QNMs and grey body factors for BHs have been found using WKB approach. Given that, the boundary conditions for QNMs and grey body factors are the same for BHs and WHs, it is reasonable to consider such similar correspondence in case of WHs too. In fact, in terms of tortoise coordinate $r_{T}$, we can distinguish  WH from a BH. $-\infty< r_{T}<+\infty$ in case of BH represents the event horizon and the one side of the asymptotic region, while $-\infty< r_{T}<+ \infty$ takes into account the two asymptotically flat space-time regions of a WH with no horizon. Though, the literature on QNMs of WHs is quite vivid, the calculation of grey body factors is case-specific.
 
 With the advent of GW astronomy, the role of QNMs and greybody factors in WH physics has gained attention \cite{Rosato:2025rgy}, \cite{gb}. These quantities help characterize the spacetime's response to perturbations and provide insights into its observable signatures. Specifically, QNMs are determined by the WH geometry and boundary conditions, whereas greybody factors represent the transmission likelihood of waves through the effective potential. Therefore, both can potentially serve as distinguishing signatures between BHs and WHs in observational contexts.
 
 Perturbations in the WH geometry such as small deviations in the background metric or test field evolutions—lead to a Schrödinger-like wave equation that governs the dynamics of the perturbation field \cite{Bolokhov:2024otn}, \cite{Murshid:2022ssj}, \cite{gb}, \cite{Iyer:1986vv}, \cite{Fernandes:2020nbq}, \cite{Lutfuoglu:2025hjy}. Thus, it is speculated that both the QNMs and grey body factors may be considered as tools to distinguish WHs from BHs in the context of GW astronomy and related observational studies.  One may perturb a WH geometry given by (\ref{eq9}), considering small deviations in the background space-time geometry or equivalently by examining the evolution of test fields in the underlying geometry. The  Schrödinger-like wave equation can be written as \cite{Bolokhov:2024otn}, \cite{gb}
 \begin{equation}
 	\left(\dfrac{d^{2}}{dr_{T}^{2}}+\omega^{2}-V(r_{T})\right)\Psi(r_{T})=0\label{eq45}
 \end{equation}
 Here, $\omega$ stands for the frequency of oscillation, $V$ is the effective potential that depends on the WH geometry and the tortoise coordinate $r_{T}$ has the expression 
 \begin{equation}
 	r_{T}(r)=\int \dfrac{dr}{e^{\Phi}\sqrt{1-\dfrac{b(r)}{r}}}
 \end{equation}  
 The explicit form of the effective potential is given by 
 \begin{equation}
 	V(r)=e^{2\Phi}\left(\dfrac{l(l+1)}{r^{2}}-\dfrac{(rb'-b)}{2r^{3}}+\dfrac{\Phi'(r)}{r}\left(1-\dfrac{b(r)}{r}\right)\right)\label{eq12}
 \end{equation} for massless scalar field and 
 \begin{equation}
 	V(r)=e^{2\Phi}\dfrac{l(l+1)}{r^{2}}
 \end{equation} for electromagnetic field.  Here $l$ (the angular momentum quantum no.) represents the multipole number. \\
 The QNMs ($Re(\omega)+i Im(\omega)$) of WHs are the solutions of the above wave equation (\ref{eq45}) with the purely outgoing boundary conditions at infinity i.e, 
 \begin{equation}
 	\Psi(r_{T})\approx \Psi_{I} e^{\pm i\omega r_{T}}, ~r_{T}\rightarrow\pm \infty
 \end{equation}
 It is to be noted that, the QNM gives information about the evolution of the perturbed field as follows: the real part of $\omega$ identifies the oscillation of the signal while the imaginary part measures the damping factor for the energy loss of gravitational radiation. More specifically, if $\Psi(r_{T})\approx \Psi_{I} e^{ i\omega r_{T}}, ~r_{T}\rightarrow\pm\infty$ then
 $Im(\omega)<0$ gives unstable perturbations with exponential growth of the perturbed field and $Im(\omega)>0$, implies stability of the perturbed field. Reversely, if $\Psi(r_{T})\approx \Psi_{I} e^{ -i\omega r_{T}}, ~r_{T}\rightarrow\pm\infty$  then $Im (\omega)<0$ gives stable perturbations and $Im(\omega)>0$ gives unstable perturbations. 
 On the other hand, due to Schrödinger like nature of wave equation (\ref{eq45}) the WKB approximation is preferable in this arena.
 
Thus, using WKB approach the QNM frequency $\omega$ near the WH throat takes the form \cite{Konoplya:2023moy}, \cite{Iyer:1986vv}, \cite{Fernandes:2020nbq}
 \begin{equation}
 	\omega=\dfrac{e^{\Phi(r_{0})}}{r_{0}}\left(l+\dfrac{1}{2}\right)-i\left(n+\dfrac{1}{2}\right)\dfrac{e^{\Phi(r_{0})}}{\sqrt{2}r_{0}}+O(l^{-1})\label{eq15}
 \end{equation}
 Hence, the real part has a relation with the shadow radius as 
 \begin{equation}
 	Re(\omega)=\dfrac{1}{r_{sh}}\left(l+\dfrac{1}{2}\right)\label{eq16}
 \end{equation}
The above expression shows that the shadow radius depends not only on the WH throat but also on the outer light ring $r_{ph}>r_{0}$. The above interrelation is exact in the eikonal limit (large $l$), while the above relation is very close for small values of $l$ as well. Also, as in the eikonal limit the electromagnetic and the scalar field behave identically so the interrelation between shadow radius and the real part of the QNMs remains same for scalar field propagation also. Now, due to smoothness of $b(r)$ and $\Phi(r)$, it is possible to have the Taylor series expansion about the WH throat at $r=r_{0}$ as

\begin{eqnarray}
	b(r)=b_{0}+b_{1}(r-r_{0})+b_{2}(r-r_{0})^{2}+...\label{eq18}\\
	\Phi(r)=\Phi_{0}+\Phi_{1}(r-r_{0})+\Phi_{2}(r-r_{0})^{2}+...\label{eq19}
\end{eqnarray}
It is worth noting that, while a prefactor is often introduced to ensure the correct behavior in asymptotically flat spacetimes, it is not necessary in the current analysis since our focus is on the vicinity of the wormhole throat. Furthermore, in the limit of large angular momentum quantum number $l$, the QNMs can be approximated analytically using the WKB method.
  The formula for the QNM frequencies to the $K-th$ order in perturbation takes the form 
 \begin{equation}
 	i\dfrac{\omega^{2}-V_{m}}{\sqrt{-2V_{m}''}}-\sum_{i=2}^{K}\Gamma_{i}=n+\dfrac{1}{2}\label{eq19*}
 \end{equation}
 Here, $V_{m}$ is the maximum of the potential , its second order derivative w.r.t the tortoise coordinate is denoted by $V_{m}''$, all higher order corrections are contained in $\Gamma_{i}'$s and $K$ stands for the order of the WKB approximation. It is important to note that larger $n$ does not give a better approximation for the quasinormal frequencies. In fact, the value of $K$ for best value of quasinormal frequency is not unique it depends on $(n,l)$. Further, one can associate the shape of WH with the imaginary part of 
 QNM as
 \begin{equation}
 	Im(\omega)=\dfrac{\sqrt{(b_{1}-1)(b_{0}\Phi_{1}-1)}}{\sqrt{2}r_{sh}}\label{eq30}
 \end{equation} Therefore, the imaginary part which governs the damping or decay of the wave amplitude over time is also characterized by the  radius of the shadow. Precisely, larger shadow radius indicates less damping. Further, $b'(r=b_{0})=1$ gives $Im(\omega)=0$ i.e, we have pure real modes as the standing waves of an oscillating string with fixed ends at the throat. 
 	In the present analysis, the condition $Im(\omega_{0})= 0$ for the fundamental mode implies the absence of exponential growth for the considered scalar perturbations, which we interpret as a sign of \emph{linear stability} within the chosen parameter space. Physically, $Im(\omega)$ controls the damping rate of perturbations. Negative values indicate decay (stable configurations), while positive values correspond to runaway growth (instability). However, a vanishing imaginary part should be viewed with caution, as it does not guarantee complete dynamical stability. In particular, gravitational (spin-2) perturbations and possible mode couplings were not addressed here. A rigorous stability criterion would require constructing the Regge--Wheeler--type master equations for tensorial perturbations in the regularized 4D EGB background and scanning for unstable modes. If such instabilities exist, they would dominate the late-time evolution, suppressing the usual ringdown interpretation and potentially triggering collapse to a BH. These aspects will be explored in future work. Nevertheless, the present results indicate that within the test-field approximation and for the explored Gauss--Bonnet parameter range, the EGB wormhole geometries do not exhibit scalar-induced dynamical instabilities, supporting their role as physically plausible configurations for observational studies.

Conversely, to analyze wave scattering near the WH, it is essential to consider how waves interact with the potential barrier surrounding the compact object. As a result of this interaction part of the wave is reflected, while the rest is transmitted through the barrier. The greybody factors quantify both of these processes, irrespective of whether the wave originates from spatial infinity or from the asymptotic region on the opposite side of the WH. Accordingly, the boundary conditions are adjusted as follows
 \begin{eqnarray}
 	\Psi=e^{-i\Omega r_{T}}+R e^{i\Omega_{r_{T}}},~r_{T}\rightarrow +\infty\nonumber\\
 	\Psi=T e^{-i\Omega r_{T}},~r_{T}\rightarrow-\infty\label{eq21}
 \end{eqnarray} with $R$ and $T$ as reflection and transmission coefficients. Here, the frequency $\Omega$ is real and continuous for scattering phenomena and is distinct from the complex and discrete QNM frequency. The transmission coefficient $T$ is usually termed as the grey body factor as it measures the fraction of the wave which traverses the potential barrier and as a result it contributes to the emission of radiation from the compact objects (BH or WH). Thus, one may write the grey body factor corresponding to the angular momentum number $l$ as
 \begin{equation}
 	\Gamma_{l}(\Omega)=|T|^{2}=1-|R|^{2}\label{eq23}
 \end{equation}
 Using WKB expression for the grey body factor one has \cite{Konoplya:2024lir}
 \begin{equation}
 	\Gamma_{l}(\Omega)=\dfrac{1}{1+e^{2\pi i k}}\label{eq24*}
 \end{equation}
 Now, the effective potential can be expanded in powers of $l$ as
 \begin{equation}
 	V(r_{T})=l^{2}V_{0}(r_{T})+lV_{1}(r_{T})+l^{-1}V_{2}(r_{T})+...\label{eq24}
 \end{equation}
 and as a result the eikonal approximation can be derived from the first order WKB approximation as \cite{Konoplya:2024lir}
 \begin{equation}
 	\Omega=l\sqrt{V_{00}}-ik\sqrt{\dfrac{-V_{00}''}{2V_{00}}}+O(l^{-1})\label{eq25}
 \end{equation}
where $V_{00}$ is the value of the function $V_{0}(r_{T})$ in its maximum and $V''_{00}$ is value of second derivative of $V_{0}(r_{T})$.  As a consequence, $k$ can be expressed as a function of the real frequency $\Omega$ \cite{Konoplya:2024lir}
 \begin{equation}
 	-ik=\dfrac{\Omega^{2}-l^{2}V_{00}}{l\sqrt{-2V_{00}}}+O(l^{-1})=-\dfrac{\Omega^{2}-Re(\omega_{0})^{2}}{4 Re(\omega_{0})Im(\omega_{0})}+O(l^{-1})\label{eq26}
 \end{equation}
 Hence, one may associate the transmission coefficient to the grey body factors $\Gamma_{l}(\Omega)$ with the fundamental mode $\omega_{0}$ as \cite{Konoplya:2024lir}
 \begin{equation}
 	\Gamma_{l}(\Omega)=|T|^{2}=\left(1+e^{^{2\pi\frac{\Omega^{2}-Re^{2}\omega_{0}}{4Re(\omega_{0})Im(\omega_{0})}}}\right)\label{eq27}
 \end{equation}
 It is to be noted that the above relation is exact in the eikonal limit $l\rightarrow \pm \infty$ and an approximate one for small $`l$'. To find the interrelation between shadow radius with grey body factor we consider equations (\ref{eq18})-(\ref{eq26}) to get
 \begin{eqnarray}
 	V_{00}=e^{2\Phi_{0}}\left(\dfrac{4l(l+1)}{2r_{0}^{4}}\left(\left(2r_{0}^{2}(\Phi_{2}+\Phi_{1}^{2})\right)-1\right)+\dfrac{b_{2}}{r_{0}^{3}}(1-2\Phi_{1}r_{0})-\dfrac{3b_{3}}{r_{0}^{2}}\right)
 \end{eqnarray}
 Choosing $\Phi_{2}+\Phi_{1}^{2}=\dfrac{1}{2r_{0}^{2}}$ and $\dfrac{3b_{3}}{r_{0}^{2}}+\dfrac{b_{2}}{r_{0}^{3}}=0$,
 we get
 \begin{equation}
 	V_{00}''=-l(l+1)b_{2}^{2}\left(\dfrac{e^{\Phi_{0}}}{r_{0}}\right)^{2}
 \end{equation}
Thus, 
\begin{eqnarray}
	Re(\omega_{0})=l\left(\dfrac{e^{\Phi_{0}}}{r_{0}}\right)\sqrt{l(l+1)}\\
	Im(\omega_{0})=\dfrac{b_{2}}{2\sqrt{2}}\\
	\Gamma_{l}(\Omega)=1+e^{\Lambda}\label{eq32}
\end{eqnarray}
where,
\begin{equation}
	\Lambda=\dfrac{\pi}{\sqrt{2}}\left(\dfrac{\Omega^{2}-\dfrac{l^{3}(l+1)}{r_{0}^{2}}e^{2\Phi_{0}}}{b_{2}\dfrac{e^{\Phi_{0}}}{r_{0}}\sqrt{l^{3}(l+1)}}\right)
	\end{equation}
Equivalently, the above expression in terms of shadow radius can be written as 
\begin{equation}
		\Lambda=\dfrac{\pi}{\sqrt{2}}\left(\dfrac{\Omega^{2}-\dfrac{l^{3}(l+1)}{r_{sh}^{2}}}{b_{2}\dfrac{1}{r_{sh}}\sqrt{l^{3}(l+1)}}\right)
\end{equation}
The expression of grey body factor in (\ref{eq32}) shows that as the damping increases (conversely the radius of the shadow decreases), there will be some value at which the shadow radius will obey the inequality $\Omega^{2}<\dfrac{l^{3}(l+1)}{r_{sh}^{2^{}}}$ thereby making $\Lambda<0$. Moreover, as $\Lambda\rightarrow -\infty$, $\Gamma_{l}\rightarrow 1$. This might be interpreted in the manner : Lesser is the shadow radius, more is the effect of damping and GBF tends to be unity. When a WH exhibits a greybody factor close to unity, it indicates that the structure is nearly transparent to incoming radiation or perturbations. The grey body factor measures the tendency that radiation will successfully penetrate the potential barrier surrounding a compact object. A grey body factor of unity signifies full transmission, meaning that no reflection occurs, and the radiation can reach the WH throat. They either emerge on the other side or continue to infinity if outgoing.  It is to be noted that as the shadow grows larger, the value of GBF obeys i.e, $\Gamma_{l}>>1$.   In standard black hole scenarios, the greybody factor $\Gamma_\ell$ is interpreted as the transmission probability of waves across the effective potential barrier, and hence it satisfies $0 \leq \Gamma_\ell \leq 1$. However, in the case of wormhole geometries considered here, the formal expression for the greybody factor is obtained by relating the transmission coefficient to the absorption cross-section normalized by the geometric cross-section. This ratio can exceed unity because the effective interaction area, influenced by the wormhole throat structure and multiple scattering effects, can be significantly larger than the naive geometric area. Therefore, values of $\Gamma_\ell > 1$ should not be interpreted as probabilities, but rather as enhancements in the effective absorption relative to the geometric limit, which arise due to the non-trivial topology and extended scattering region of the wormhole spacetime.

\section{TWHs in EGB Theory: QNMs and GBFs}
\subsection{TWHs in EGB theory}
	Action for the D-dimensional EGB gravity theory can be written as \cite{Fernandes:2020nbq}-\cite{Ghosh:2020syx}
\begin{equation}
	\mathcal{A}=\dfrac{1}{16 \Pi}\int d^{D}x~\sqrt{-g}\left(R+\dfrac{\alpha}{D-4}\mathcal{L}_{GB}\right)+S_{matter}\label{eq1}
\end{equation}
Here, $S_{matter}$ is the matter Lagrangian of the system , $R$ is the Ricci scalar and the expression for $\mathcal{L}_{GB}$ is the Lagrange density function for EGB gravity has the expression
\begin{equation}
	\mathcal{L}_{GB}=R^{abcd}~R_{abcd}-4R^{ab}R_{ab}+R^{2}\label{eq2}
\end{equation}
Now, if we vary the action integral with respect to the metric $g_{\mu\nu}$ then we get the modified Einstein field equations as \cite{Ghosh:2020vpc}
\begin{equation}
	G_{ab}+\dfrac{\alpha}{D-4}H_{ab}=8\Pi T_{ab}\label{eq3}
\end{equation}
Here $G_{ab}$ is the Einstein tensor (i.e, $G_{ab}=R_{ab}-\dfrac{1}{2}R g_{ab}$), the explicit form of the second rank Lancoz tensor $H_{ab}$ is 
\begin{equation}
	H_{ab}=2\left(R~R_{ab}-2R_{ac}~R^{c}_{b}-2R_{acbd}~R^{cd}-R_{acde}~R^{cde}_{b}\right)-\dfrac{1}{2}\mathcal{L}_{GB}~g_{ab}\label{eq4}
\end{equation}
and the energy momentum tensor $T_{ab}$ is obtained from the matter action as 
\begin{equation}
	T_{ab}=-\dfrac{2}{\sqrt{-g}}\dfrac{\delta (\sqrt{-g}~S_{m})}{\delta g^{ab}}\label{eq5}
\end{equation}
As mentioned earlier, the GB term becomes a total derivative term in 4D space-time, resulting no effect in 4D field equations so by re-scaling the coupling constant by $\dfrac{\alpha}{D-4}$ one gets \cite{Chakraborty:2025cpp}
\begin{equation}
	\dfrac{g_{ac}}{\sqrt{-g}}\dfrac{\delta \mathcal{L}_{GB}}{\partial g_{bc}}=\dfrac{\alpha(D-2)(D-3)}{2(D-1)}\kappa^{2}\delta^{a}_{b}~, \label{eq6}
\end{equation} corresponding to a maximally symmetric space-times with curvature scale $\kappa$ and hence the variation of the GB term in the action does not vanish \cite{Cognola:2013fva}. Thus, the 4D spherical solution with the regularization process will be used in the next subsection as the basic equations for WH geometry \cite{Cognola:2013fva}. It is to be noted that the above spherical solution is identical to that obtained with other regularization technique \cite{Lu:2020iav}. Due to inhomogeneous nature of the WH geometry it is reasonable to consider anisotropic fluid as the matter source having energy-momentum tensor
\begin{equation}
T^{a}_{b}=(\rho+p_{t})u_{a}u^{b}+p_{t}\delta_{a}^{b}+(p_{r}-p_{t})v_{a}v^{b}\label{eq9}
\end{equation}
Here, $u^{a}$ is the $n$- velocity vector, $v^{a}$ is the unit space-like vector in the radial direction, $\rho(r)$ is the energy density, $p_{r}$ is the radial pressure and $p_{t}$ is the transverse pressure. Now using, the space-time geometry (\ref{eq9*}) and the matter component (\ref{eq9}), the explicit form of the Einstein field equations (\ref{eq3}) in the limit $D\rightarrow 4$ has the explicit form as \cite{Jusufi:2020yus}
\begin{equation}
8\Pi \rho(r)=\dfrac{\alpha b(r)}{r^{6}}(2rb'(r)-3b(r))+\dfrac{b'(r)}{r^{2}}\label{eq10}
\end{equation}
\begin{equation}
8\Pi p_{r}(r)=\dfrac{\alpha b(r)}{r^{6}}(4\phi'r(r-b(r))+b(r))+\dfrac{2\phi '(r-b(r))}{r^{2}}-\dfrac{b(r)}{r^{3}}\label{eq11}
\end{equation}
\scriptsize
\begin{eqnarray}
8\Pi p_{t}(r)=\left(1-\dfrac{b(r)}{r}\right)\left((\phi''+\phi'^{2})\left(1+\dfrac{4\alpha b(r)}{r^{3}}\right)+
\dfrac{1}{r}\left(\phi'-\dfrac{(rb'(r)-b(r))}{2r(r-b(r))}\right)\left(1-\dfrac{2\alpha b(r)}{r^{3}}\right)-\dfrac{(rb'(r)-b(r))}{2r(r-b(r))}\phi'\left(1-\dfrac{8\alpha}{r^{2}}+\dfrac{12\alpha b(r)}{r^{3}}\right)\right)-\nonumber\\
\dfrac{2\alpha b^{2}(r)}{r^{6}}
\end{eqnarray}
\normalsize
where $'$ stands for derivative w.r.t $r$.
We now pick two solutions from the paper \cite{Chakraborty:2025cpp} to find the QNM spectra and GBFs namely an isotropic solution and an anisotropic solution involving the gauss bonnet parameter $\alpha$.
\subsection{QNM and Grey body spectra for isotropic and anisotropic EGB WHs}

\begin{figure}[h!]
	\begin{minipage}{0.3\textwidth}
		\includegraphics[height=5.5cm,width=7.5cm]{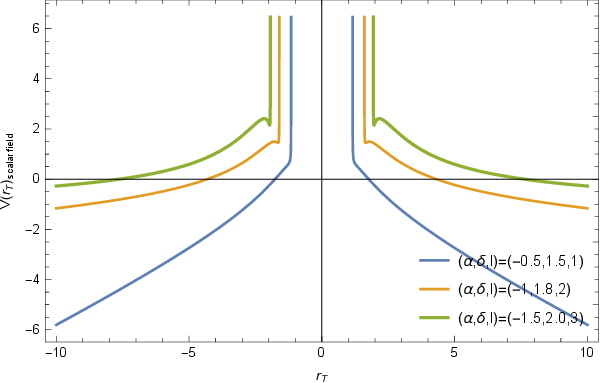}
	\end{minipage}~~~~~~~~~~~~~~~~~~~~~~~~~~~~~~~~~~~~~~~~~~
	\begin{minipage}{0.3\textwidth}
		\includegraphics[height=5.5cm,width=7.5cm]{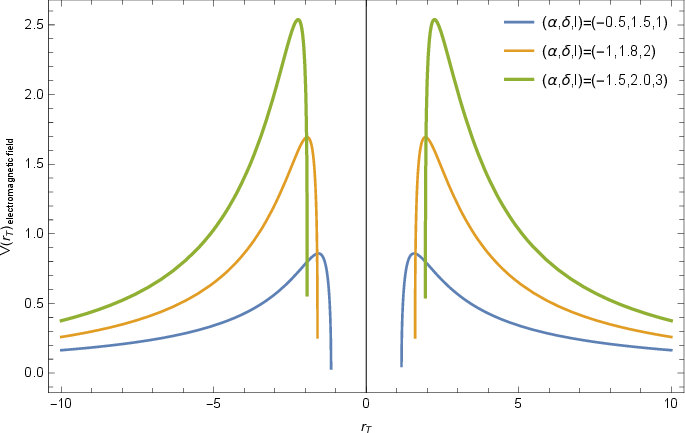}
	\end{minipage}
	\caption{Variation of potential with tortoise coordinate for massless scalar field (left) and electromagnetic field (right) in case of anisotropic WH solution}\label{F6}
\end{figure}

\begin{figure}[h!]
	\begin{minipage}{0.3\textwidth}
		\includegraphics[height=5.5cm,width=7.5cm]{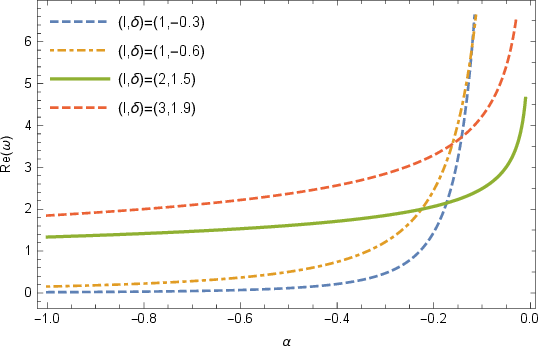}
	\end{minipage}~~~~~~~~~~~~~~~~~~~~~~~~~~~~~~~~~~~~~~~~~~
	\begin{minipage}{0.3\textwidth}
		\includegraphics[height=5.5cm,width=7.5cm]{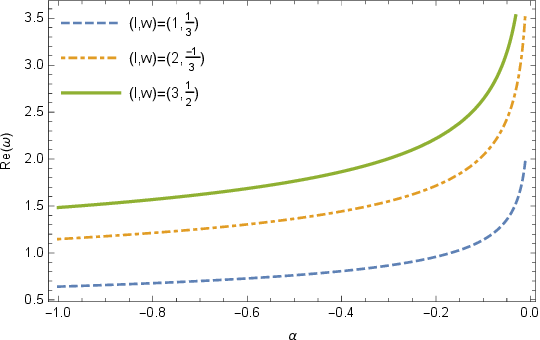}
	\end{minipage}
	\caption{Variation of QNM spectra (arbitrary mode) with Gauss Bonnet parameter ($\alpha$) for anisotropic (left) and isotropic (right with ($\rho_{0},~\Lambda=0.1$ chosen)) WH solutions}\label{F3}
\end{figure}

\begin{figure}[h!]
	\begin{minipage}{0.3\textwidth}
		\includegraphics[height=5.5cm,width=7.5cm]{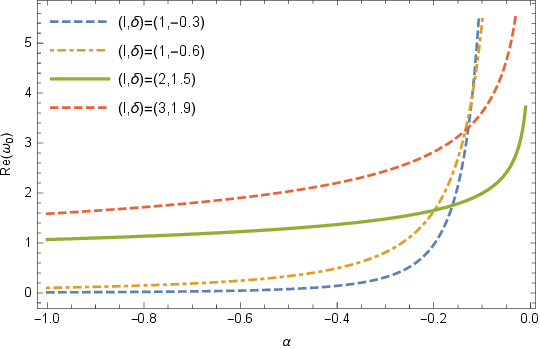}
	\end{minipage}~~~~~~~~~~~~~~~~~~~~~~~~~~~~~~~~~~~~~~~~~~
	\begin{minipage}{0.3\textwidth}
		\includegraphics[height=5.5cm,width=7.5cm]{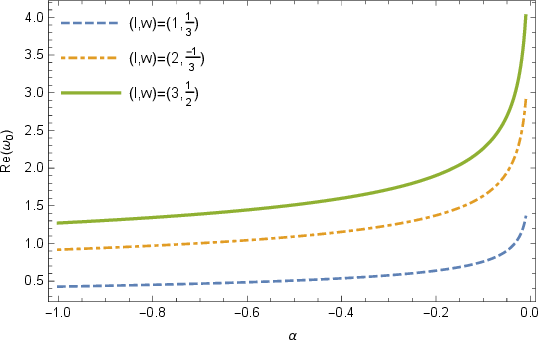}
	\end{minipage}
\caption{Variation of fundamental mode of QNM spectra  with Gauss Bonnet parameter ($\alpha$) for anisotropic (left) and isotropic (right with ($\rho_{0},~\Lambda=0.1$ chosen)) WH solutions}\label{F2}
\end{figure}

\begin{figure}[h!]
\centering	\begin{minipage}{0.3\textwidth}
		\includegraphics[height=5.5cm,width=8.5cm]{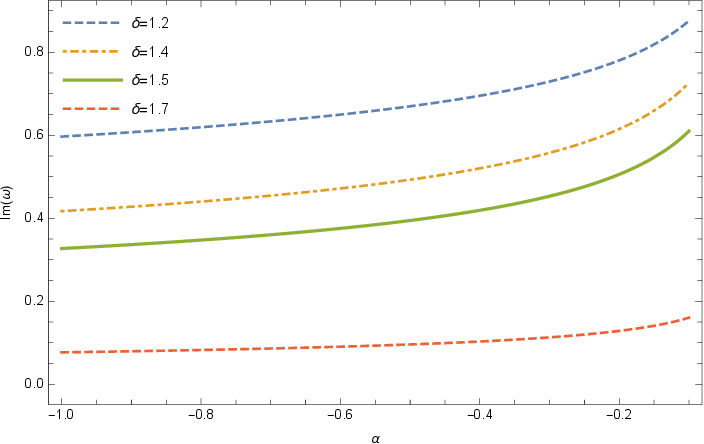}
	\end{minipage}
\caption{Variation of Im($\omega$) with the gauss bonnet parameter $\alpha$ for the anisotropic EGB WH solution and $Im(\omega_{0})=0$}\label{F5}
\end{figure}

\textbf{Anisotropy Case:} The shape function is given by \cite{Chakraborty:2025cpp}
\begin{equation}
	b(r)=-\dfrac{r^{3}}{4\alpha},~\alpha<0
\end{equation}  and the red-shift function is given by 
\begin{equation}
\phi(r)=\dfrac{10-3\delta}{16\delta}\ln \left(r^{2}+\dfrac{(1+2\delta)\alpha}{\delta}\right)
\end{equation}
The radius of photon ring or photon sphere in this case is $r_{ph}=2\sqrt{-\alpha}$ and the radius of the shadow is given by
\begin{equation}
	r_{sh}=2\sqrt{-\alpha}\left(\dfrac{\alpha}{\delta}(1-2\delta)\right)^{\dfrac{3\delta-10}{16\delta}}
\end{equation}
Thus we have, 
\begin{eqnarray}
	V(r_{T})_{scalar~field}=\left(r_{T}^{2}+\dfrac{(1+2\delta)\alpha}{\delta}\right)^{\dfrac{10-3\delta}{8\delta}}\left(\dfrac{l(l+1)}{r_{T}^{2}}+\dfrac{1}{4\alpha}+\dfrac{(10-3\delta)\left(1+\dfrac{r_{T}^{2}}{4\alpha}\right)}{8\delta\left(r_{T}^{2}+\frac{(1+2\delta)\alpha}{\delta}\right)}\right)\label{eq48}\\
	V(r_{T})_{electromagnetic~field}=\left(\dfrac{l(l+1)}{r_{T}^{2}}\right)\left(r_{T}^{2}+\dfrac{(1+2\delta)\alpha}{\delta}\right)^{\dfrac{10-3\delta}{8\delta}}\label{eq49}\\
	Re(\omega)=\dfrac{l+\frac{1}{2}}{2\sqrt{-\alpha}\left(\dfrac{\alpha}{\delta}(1-2\delta)\right)^{\dfrac{3\delta-10}{16\delta}}}\\
	Re(\omega_{0})=\dfrac{l}{2\sqrt{-\alpha}\left(\dfrac{\alpha}{\delta}(1-2\delta)\right)^{\dfrac{3\delta-10}{16\delta}}}\\
	Im(\omega)=\dfrac{\sqrt{\left(2\sqrt{-\alpha}\left(\dfrac{(10-3\delta)\sqrt{-\alpha}}{4\alpha(1-2\delta)}\right)\right)-1}}{2\sqrt{-\alpha}\left(\frac{\alpha}{\delta}(1-2\delta)\right)^{\dfrac{3\delta-10}{16\delta}}},~~	Im(\omega_{0})=0
\end{eqnarray}
The variations of potential barrier (for massless scalar field and electromagnetic field) as a function of the tortoise coordinate have been shown graphically in FIG. (\ref{F6}). 
The plots showing the variation of QNM spectra (both real and imaginary part) are shown in FIGs (\ref{F3})  (left), (\ref{F2})  (left) and (\ref{F5}) respectively.

\begin{figure}[h!]
	\begin{minipage}{0.3\textwidth}
	\includegraphics[height=5.5cm,width=7.5cm]{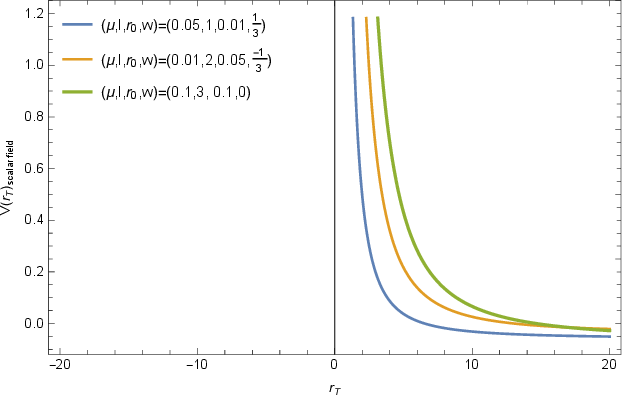}
	\end{minipage}~~~~~~~~~~~~~~~~~~~~~~~~~~~~~~~~~~~~~~~~~~
	\begin{minipage}{0.3\textwidth}
		\includegraphics[height=5.5cm,width=7.5cm]{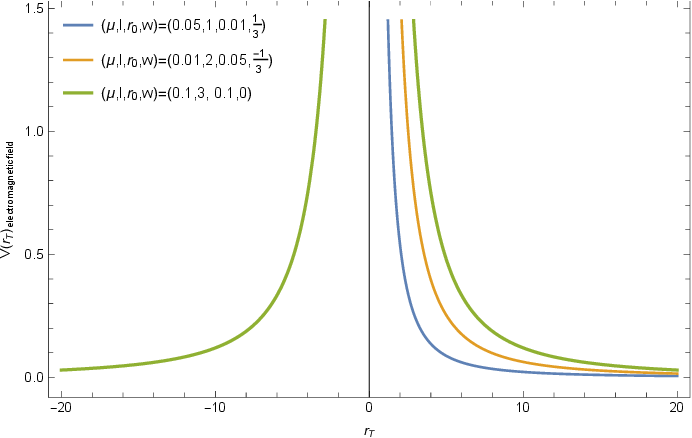}
\end{minipage}
	\caption{Variation of potential with tortoise coordinate for massless scalar field (left) and electromagnetic field (right) for isotropic WH solution ($\rho_{0},~\Lambda=0.1$ chosen)}\label{F7}
\end{figure}

\textbf{Isotropy Case:} The shape function has the expression \cite{Chakraborty:2025cpp}
\begin{equation}
	b(r)=r_{0}+\dfrac{\rho_{0}r_{0}^{3}}{\mu+3}\left(\left(\dfrac{r}{r_{0}}\right)^{\mu+3}-1\right)
\end{equation} (the positive branch of solution as $\alpha\rightarrow 0$) and if we define $\rho$ as
\begin{equation}
	\rho=\rho_{0}\left(\left(\dfrac{r}{r_{0}}\right)^{-\mu}+\Lambda\right)
\end{equation} where $\mu, ~\Lambda>0$ then,
\begin{equation}
	\exp(\phi(r))=\left(\left(\dfrac{r}{r_{0}}\right)^{-\mu}+\Lambda\right)^{\dfrac{-w}{w+1}}\label{eq55}
\end{equation} $w$ being the EoS parameter. 
The expressions for potential barrier in case of  both massless scalar field and electromagnetic field are given by 
\begin{equation}
	V(r_{T})_{scalar~field}=\left(\left(\dfrac{r_{T}}{r_{0}}\right)^{-\mu}+\Lambda\right)^{\dfrac{-2w}{w+1}}\left(\dfrac{l(l+1)}{r_{T}^{2}}-A+B\times C\right)\label{eq56}
	\end{equation}
where
\begin{equation}	A=\dfrac{\rho_{0}}{2}\left(\dfrac{r_{T}}{r_{0}}\right)^{\mu}\left(\dfrac{\mu+2}{\mu+3}\right)+\dfrac{\rho_{0}}{2(\mu+3)}\left(\dfrac{r_{T}}{r_{0}}\right)^{-3}-\dfrac{r_{0}}{2r_{T}^{3}}
	\end{equation}
\begin{equation}
	B=\dfrac{w\mu}{r_{0}r_{T}(w+1)}\times \dfrac{\left(\dfrac{r_{T}}{r_{0}}\right)^{-\mu-1}}{\left(\left(\dfrac{r_{T}}{r_{0}}\right)^{-\mu}+\Lambda\right)}
	\end{equation}
\begin{eqnarray}
	C=1-\dfrac{r_{0}}{r_{T}}-\dfrac{\rho_{0}r_{0}^{3}}{(\mu+3)r_{T}}\left(\left(\dfrac{r_{T}}{r_{0}}\right)^{\mu+3}-1\right)\\
		V(r_{T})_{electromagnetic~field}=\dfrac{l(l+1)}{r_{T}^{2}}\left(\left(\dfrac{r_{T}}{r_{0}}\right)^{-\mu}+\Lambda\right)^{\dfrac{-2w}{w+1}}\label{eq60}
\end{eqnarray}
The plots showing the variation of potential barrier (both for massless scalar field and electromagnetic field) w.r.t the tortoise coordinate are given in FIG. (\ref{F7}).
In both the isotropic as well as anisotropic cases, we find the potential barrier is occurring near $r_{T}=0$. In the anisotropic case, more negative is the Gauss Bonnet parameter the peaks of the potential are higher. However, in both the cases driven by isotropy and anisotropy we get symmetric potential barrier curves w.r.t $r_{T}=0$ hinting that the geometry of the spacetime is symmetric around the throat of the EGB WHs.

 The radius of photon sphere takes the form
 \begin{equation} r_{ph}=\left(\dfrac{-3\alpha}{\rho_{0}(1+\Lambda)}\right)^{\frac{1}{4}}
 	\end{equation}
 	 and radius of the shadow is given by
 	 \begin{equation} r_{sh}=\left(\dfrac{-3\alpha}{\rho_{0}(1+\Lambda)}\right)^{\frac{1}{4}}(1+\Lambda)^{\frac{w}{w+1}}
 	 	\end{equation}
 	 	It is to be noted that, since the shape function corresponds to $\alpha\rightarrow0$ and we find that $r_{sh}\rightarrow 0$ as $\alpha\rightarrow 0$ hence one has $Im(\omega)\rightarrow\infty$ (infinite damping as gauss bonnet parameter becomes vanishing). But, imaginary part of the fundamental mode comes out to be zero i.e, $Im(\omega_{0})=0$. Moreover, the expressions for real part of QNM frequency (both arbitrary and fundamental) are given by 
 \begin{eqnarray}
 	Re(\omega)=\dfrac{l+\frac{1}{2}}{\left(\dfrac{-3\alpha}{\rho_{0}(1+\Lambda)}\right)^{\frac{1}{4}}(1+\Lambda)^{\frac{w}{w+1}}}\\
 	Re(\omega_{0})=\dfrac{l}{\left(\dfrac{-3\alpha}{\rho_{0}(1+\Lambda)}\right)^{\frac{1}{4}}(1+\Lambda)^{\frac{w}{w+1}}}
 \end{eqnarray}
The plots (right) showing variation of QNM spectra with $\alpha$ are given in FIGs (\ref{F3}) (arbitrary mode) and (\ref{F2}) (fundamental mode) respectively.
The expression for GBF is given by equation (\ref{eq27}). In the present model, we have $Im(\omega_{0})=0$. Thus, $\Gamma_{l}\rightarrow 1$ if $\Omega^{2}<Re^{2}(\omega_{0})$, meaning in order to have a finite (close to unity) value of grey body factor, the QNM frequencies should fall below the fundamental mode $\Omega_{0}$. If there exists a frequency that falls above $\Omega_{0}$, then $\Gamma_{l}$ corresponding to that frequency will be infinite (indicating infinite transmission). Thus, the QNM frequency $\Omega_{0}=Re(\omega_{0})+i Im(\omega_{0})$ acts as a critical frequency/ threshold frequency which in the present model reduces to $Re(\omega_{0})$ only as damping effect via $Im(\omega_{0})$ is vanishing. Further, since $Im(\omega_{0})=0$ thus the WH configurations are stable.

\subsection{Some important remarks}
\subsubsection{NEC Violation and Its Impact on Spectra:}
For the isotropic  ($p_{r}=p_{t}=p$) EGB WH configuration, the NEC at the throat can be written as
\[
8\pi(\rho+p)\big|_{r_0} = \frac{\rho_0 r_0^{2}-1}{r_0^{2}}\left(1+\frac{2\alpha}{r_0^{2}}\right),
\]
which typically becomes negative when the flare-out condition $b'(r_0)<1$ holds and $\alpha>-r_0^{2}/2$. Thus, a certain degree of NEC violation persists in this case, modulated by the Gauss--Bonnet coupling. On the contrary, for the anisotropic WH defined by $b(r)=-r^{3}/(4\alpha)$ with $\alpha<0$, the throat conditions give
\[
8\pi(\rho+p_r)\big|_{r_0} = \frac{1}{r_0^{2}}, \qquad
8\pi(\rho+p_t)\big|_{r_0} = \frac{11}{4r_0^{2}},
\]
indicating that the matter sector does not violate the NEC at the throat. The effective flare-out is supported by the higher-curvature terms. Importantly, the QNM and gravitational bound-state spectra depend solely on the background geometry $(b,\Phi;\alpha)$ and are therefore insensitive to the degree/ quantitative measure of NEC violation. Any variation in NEC affects the spectra only through corresponding changes in the metric functions.

\subsubsection{Nature of Perturbations and QNM Spectra: }
 In this study, we have primarily analyzed QNMs by considering massless scalar field perturbations, as shown in the effective potentials given by equation (\ref{eq48}) for the anisotropic case and equation (\ref{eq56}) for the isotropic case. For completeness, we have also provided the corresponding electromagnetic potentials in equation (\ref{eq49}) and equation (\ref{eq60}). It is important to note that the QNM spectra depend on the spin of the perturbing field. For electromagnetic ($s=1$) and gravitational ($s=2$) perturbations, the effective potential includes additional spin-dependent terms compared to the scalar case ($s=0$), which generally increase the height of the potential barrier. This results in higher real parts of the frequency (faster oscillations) and more negative imaginary parts (stronger damping) for the same multipole number $l$.
Despite these differences, the qualitative behavior such as the eikonal correspondence between the real part of the QNM frequency and the photon-sphere radius, expressed in equation (\ref{eq16}) remains valid across different spins. The choice of scalar perturbations in this work is motivated by analytical simplicity and also since scalar fields capture the essential features needed to establish the link between QNMs, WH shadows, and greybody factors. A detailed extension to electromagnetic and gravitational perturbations, especially for EGB WHs is left for future work considering its astrophysical relevance for GW observations.

\subsubsection{Greybody Factor Calculation and Sensitivity:} 
In this work, the greybody factors are computed using the WKB approximation, which is well suited for analyzing scattering in space-times with smooth potential barriers. Specifically, the first-order WKB formula for the transmission coefficient (see equation (\ref{eq24*})) was employed, and its relation to the fundamental QNM frequency was established through equation (\ref{eq27}). This approach approximates the solution of the Schrödinger-like wave equation in equation (\ref{eq45}) by matching the asymptotic WKB expansions across the turning points near the peak of the effective potential. The WKB method is preferred over direct numerical integration because it provides an analytical expression for the transmission coefficient in terms of the potential’s maximum and its derivatives, making it particularly effective in the eikonal regime (large multipole number $l$). For lower $l$, the approximation becomes less accurate, but the qualitative behavior remains consistent with numerical studies in similar geometries. In case of WHs, the tortoise coordinate covers two asymptotically flat regions without a horizon, unlike BHs where near-horizon expansions dominate. Here, the WKB method ensures that the boundary conditions for purely outgoing waves in both asymptotic regions (equation (\ref{eq21})) are satisfied by construction. Variations in the matching point near the potential peak affect the numerical value of the GBF only marginally because the exponential dependence on the barrier shape dominates the transmission coefficient. Higher-order WKB corrections can reduce residual errors, but for the purpose of identifying trends and establishing relations with shadow radius and QNM frequencies, the first-order approximation is sufficient.
\section{Justification of the QNM–shadow correspondence: A semi-analytic approach}
The correspondence between the real part of the quasinormal frequency and the wormhole shadow radius follows from two simple facts: (i) in the high-multipole (eikonal) limit the wave equation reduces to a geometric-optics problem, and (ii) the dominant contribution to the oscillation frequency in that limit is set by properties of the unstable circular null geodesic (the photon ring). Small perturbations (scalar, electromagnetic or gravitational) when written as a single master wave equation are governed by a Schrödinger-type equation (\ref{eq45}) with an effective potential whose leading angular dependence scales like $l^{2}$. Schematically, we write 
\begin{equation}
	V(r_{T})=l^{2}V_{0}(r_{T})+O(l)\label{eq65}
\end{equation}
where $r_{T}$
is the tortoise coordinate and $l$ the multipole number. In the eikonal regime $l>>1$, the $l^{2}$ term dominates and the problem becomes a short-wavelength (WKB) problem. The function $V_{0}(r_{T})$ is directly related to the geometry felt by null rays. Its maximum occurs at the radius corresponding to the unstable circular null geodesic, $r_{ph}$. Physically, the barrier that controls scattering and ringing is the same structure that defines the photon ring. At leading order the WKB condition gives,
\begin{equation}
	\omega^{2}\approx l^{2} V_{0}(r_{ph})\implies Re(\omega)\approx l \sqrt{V_{0}(r_{ph})}
\end{equation}
The metric functions enter via $V_{0}$. For a static spherically symmetric WH metric given by equation (\ref{eq9*}) one finds 
\begin{equation}
	\sqrt{V_{0}(r_{ph})}=\Omega_{ph}
\end{equation} by comparing the geometric optics dispersion for null circular geodesics and the effective potential. Here $\Omega_{ph}$ is the angular velocity of the circular null geodesic measured at infinity given by 
\begin{equation}
	\Omega_{ph}=\dfrac{e^{\Phi(r_{ph})}}{r_{ph}}
\end{equation}
These altogether lead to the equations 
\begin{equation}
	Re(\omega)\approx l \Omega_{ph}=l\dfrac{e^{\Phi(r_{ph})}}{r_{ph}}
\end{equation}
On the other hand, the shadow radius $r_{sh}$ for a distant observer is related to the photon ring by $r_{sh}=r_{ph}e^{-\Phi(r_{ph})}$. If we eliminate $\Phi$ and $r_{ph}$ we get the compact eikonal relation
\begin{equation}
	Re(\omega)\approx \dfrac{l}{r_{sh}}\approx\dfrac{\left(l+\dfrac{1}{2}\right)}{r_{sh}}
\end{equation} where $+\dfrac{1}{2}$ term is the WKB correction often included for better accuracy at finite $l$. It is to be noted that, $Re(\omega)\approx l \Omega_{ph}$  is spin independent. The derivation uses only the metric functions $\Phi(r)$ the radial metric combination that determine the null geodesics. Both isotropic and anisotropic solutions change the location $r_{ph}$ and the value of $\Phi(r_{ph})$ (hence $r_{sh}$ and $\Omega_{ph}$) but the structure of the argument is identical. Therefore the correspondence holds in both cases but the numerical values change through the metric functions, not the logic of the relation.
\section{Conclusion}
The present analysis not only strengthens the theoretical foundation of TWH physics but also opens multiple directions for future exploration. With the rapid progress of GW astronomy and the EHT imaging, the distinctive QNM spectra and greybody factors derived here could guide observational strategies for identifying potential wormhole candidates. In addition, exploring the role of modified theories such as Einstein–Gauss–Bonnet gravity provides an effective testbed for probing beyond-Einstein gravity regimes and could help constrain fundamental parameters of high-curvature corrections. Further studies could extend the framework to rotating or charged wormholes, incorporate backreaction effects, and test stability under astrophysically realistic perturbations. Ultimately, if observational signatures consistent with wormhole predictions are detected, they would not only revolutionize our understanding of compact objects but also provide insights into quantum gravity, exotic matter distributions, and the possible existence of non-trivial topological structures in our universe.
 The paper represents an exhaustive analysis of QNMs and grey body factors and how these offer a promising avenue in identifying TWHs in astrophysical observations. The relationship between shadows and quasinormal modes (QNMs) of a compact object, particularly wormholes, can be established in the way these objects interact with light and perturbations in the surrounding space-time. Quasinormal modes and greybody factors depend on how perturbations propagate along geodesics in the curved wormhole space-time. While black holes and wormholes exhibit similar perturbation responses, subtle differences in their QNM spectra and grey body factors could serve as key discriminators. Wormholes can mimic black hole shadows but have different QNM spectra due to the absence of a true horizon. This is the reason why, study of QNMs near a compact object can enable us to distinguish them. Further, there is a nice analogy between QNMs and shadow of wormhole. The shadow size is inversely related to the QNM frequency $\Omega$, linking directly to the shadow’s boundary. Moreover, the imaginary part is also related inversely to the shadow radius thereby linking it with damping effects. The dominant QNM frequencies are closely related to the photon sphere’s properties thereby getting associated with radius of the shadow.
 In wormholes, the effective photon potential determines both the shadow boundary and the ringdown frequencies of perturbations thereby linking these two key ideas. Moreover, grey body factor tends to unity as the shadow radius is very small, while $\Gamma_{l}>>1$ for large shadow radius. Thus, the paper showcases an unification of these two observational signatures in explaining a feasible WH configuration and discusses the nature of QNM spectra w.r.t the gauss bonnet parameter $\alpha$ in case of TWHs (both isotropic as well as anisotrpic) in EGB theory of gravity. It is observed that, as $\alpha\rightarrow 0$ the real part of QNM frequency (both arbitrary as well as fundamental) grows infinitely large. It is also noticed that, for large range of negative $\alpha$, $Re(\omega)/Re(\omega_{0})$ remains constant. While, $Im(\omega_{0})=0$ implies that there is no effect of damping. This makes the GBF tend to unity provided  the QNM frequencies fall below the fundamental mode $\Omega_{0}$. However, for arbitrary mode $\omega$, $Im(\omega)$ is not necessarily zero and it is shown that the damping effects are less ($Im(\omega)<1$) if $\alpha<0$. Thus, negative signature of gauss bonnet parameter leads to stable EGB TWH configurations. 
 
While the QNM–shadow relation is known for BHs and some WH models, our work derives explicit analytical expressions for isotropic and anisotropic EGB WHs. To our knowledge, this is the first systematic derivation of such relations in the context of regularized 4D EGB gravity, extending previous results restricted to GR WHs or higher-dimensional Gauss–Bonnet BHs. Further, we identify a threshold frequency condition linked to the fundamental QNM, which determines whether the greybody factor tends to unity (full transmission) or diverges (infinite transmission). This result provides a new criterion to characterize stability and transparency of anisotropic EGB WHs. Our analysis shows that for negative Gauss–Bonnet parameter $\alpha$, the imaginary part of the fundamental QNM vanishes $(Im(\omega_{0})=0)$, implying stable configurations. This complements existing studies that primarily emphasize isotropic solutions. The anisotropic case thus reveals a richer structure in a way that it demonstrates how anisotropy affects damping rates and greybody spectra. This might be crucial in discriminating WH signals from BH ringdowns in GW observations. In summary, our novelty lies in extending the QNM–shadow correspondence and greybody factor analysis into the EGB WH sector with explicit isotropic vs. anisotropic comparisons and highlighting how these features may serve as observational signatures of TWHs. From an observational perspective, several concrete signatures could help distinguish WHs from BHs. These include the appearance of narrow ringdown windows in the GW spectrum, deviations in the late-time tail compared to standard BH damping, and the possible presence of echoes or time delays arising from wave reflections between the two asymptotic regions of a WH. Similarly, greybody factors close to unity would suggest near-transparency to incoming radiation, producing anomalously high transmission compared to BH counterparts. Together, these features provide testable predictions that can be sought in high-precision GW data from detectors such as LIGO, VIRGO, KAGRA, and future observatories like LISA or the Einstein Telescope.	Regarding astrophysical feasibility, while the need for exotic matter remains a challenge, the incorporation of higher-curvature corrections via Einstein–Gauss–Bonnet gravity enhances stability and offers more realistic configurations than those in pure Einstein gravity. The isotropic and anisotropic EGB wormholes studied here suggest that stable wormhole geometries are not entirely ruled out in high-curvature regimes, especially under negative Gauss–Bonnet coupling. Therefore, experimental prospects lie in searching for anomalous ringdown signals, deviations in greybody-modulated radiation spectra and discrepancies in shadow–QNM correspondences that would not fit standard BH models. To quantify the influence of the Gauss--Bonnet parameter, we evaluated the fundamental mode ($l = 2$, $n = 0$) for representative parameter choices (isotropic: $\rho_0 = 0.1$, $\Lambda = 0.1$, $w = 1/3$). For the isotropic family, the real part of the frequency decreases from $\mathrm{Re}(\omega) \approx 3.38$ at $\alpha = -0.01$ to $\mathrm{Re}(\omega) \approx 0.60$ at $\alpha = -10$, consistent with the leading-order scaling $\mathrm{Re}(\omega) \propto |\alpha|^{-1/4}$ (follows from the eikonal relation). The anisotropic branch exhibits a stronger dependence, with $\mathrm{Re}(\omega)$ falling from $12.50$ to $0.39$ across the same range which is again in agreement with the $|\alpha|^{-1/2}$ trend for anisotropic case. For both cases, the imaginary part for the fundamental mode remains zero in the regimes considered, indicating stability. Compared to the Schwarzschild baseline ($\mathrm{Re}(\omega) \approx 0.385$ for $M = 1$), EGB WH frequencies are significantly higher for small $|\alpha|$, but approach GR-like values as $|\alpha|$ becomes large. Across the parameter range, the variation is monotonic (as clear from the plots of $Re(\omega)/Im(\omega)$ vs $\alpha$), although the anisotropic expressions contain additional $\alpha$-dependent prefactors that could induce mild non-monotonicity for certain choices of the anisotropy factor $\delta$. This systematic redshift of QNM frequencies with increasing $|\alpha|$ highlights the potential observability of Gauss--Bonnet corrections through precise ringdown measurements. The QNM frequency shifts induced by the Gauss--Bonnet coupling are significant across most of the parameter space explored, with deviations ranging from 10-100 $\%$ relative to the Schwarzschild baseline for moderate and small $|\alpha|$. Such large shifts would be clearly distinguishable with current LIGO / VIRGO sensitivity in high-SNR events. For very large $|\alpha|$, the shifts reduce to a few percent, which are beyond the precision of present detectors but fall within the expected capabilities of next-generation observatories such as the Einstein Telescope, Cosmic Explorer, and LISA. A detailed detectability analysis will require a dedicated study including waveform systematics, which we will take up in a future work. In this work we have analysed quasinormal modes and grey-body factors for massless scalar (and gave the corresponding electromagnetic) test-field perturbations of the isotropic and anisotropic EGB wormhole families. For the representative parameter sets shown in Secs.~3.2–3.3 the fundamental mode exhibits a vanishing imaginary part, $(Im(\omega_{0})=0)$, indicating linear stability of that mode within the explored parameter window. However, we stress that a complete linear stability study, in particular of gravitational (spin-2) perturbations those  have not been carried out here. Instabilities in the gravitational sector (or slowly growing modes revealed by time-domain evolution) would dominate late-time dynamics, spoil the usual damped-ringdown interpretation of the discrete QNM spectrum, modify grey-body transmission properties, and could even trigger non-linear evolution (e.g. collapse to a black hole) that would render the test-field spectra inapplicable. A full assessment therefore requires constructing the appropriate equations for tensor perturbations in the regularized 4D EGB background, scanning the model parameter space for unstable modes, and performing time-domain evolutions; we leave these tasks to future work.
 As a result theoretical progress in stability analysis, coupled with advancements in GW astronomy, enhances the prospects for observationally distinguishing wormholes from other compact objects. If detected, such structures would not only revolutionize our understanding of space-time topology but also provide deeper insights into quantum gravity and the fundamental nature of the universe or precisely will enable us to understand the universal dynamics and above all will answer the question  ``Wormholes : Myth or Reality?"

	\section*{Acknowledgment}
The authors thank the anonymous learned referees whose valuable and insightful suggestions improved the quality of the manuscript. M.C thanks Department of Mathematics, Techno India University, West Bengal and S.C thanks Department of Mathematics, Brainware University, West Bengal for providing research facilities.  M.C thanks Inter University Center for Astronomy and Astrophysics (IUCAA), Pune, India for their lecture series on `General Relativity and Quantum Field Theory' under General Relativity Refresher course 2025. S.C thanks IUCAA, Pune, India for their Visiting Associateship program.

\appendix
\section*{Appendix}
\label{app:Imw_derivation}

In this appendix we present the detailed steps that lead from the Schrödinger-like wave equation and the WKB quantization condition to equation (\ref{eq30}),
\[
Im(\omega)=\frac{1}{\sqrt{2}\,r_{ sh}}\sqrt{(b_1-1)(b_0\Phi_1-1)} .
\tag{20}
\]
\vspace{6pt}
\noindent\textbf{Step 1. Wave equation and potential:}
The radial perturbation satisfies
\[
\frac{d^2\Psi}{dr_T^2} + \big(\omega^2 - V(r)\big)\Psi = 0,
\]
with tortoise coordinate
\[
r_T(r)=\int \frac{e^{\Phi(r)}}{\sqrt{1-b(r)/r}}\;dr,
\]
and (scalar) effective potential (\ref{eq12})
\[
V(r)=e^{2\Phi(r)}\!\left(\frac{\ell(\ell+1)}{r^2}-\frac{r b'(r)-b(r)}{2r^3}+\frac{\Phi'(r)}{r}\Big(1-\frac{b(r)}{r}\Big)\right).
\]
\vspace{6pt}
\noindent\textbf{Step 2. Taylor expansions near the throat:}
Expand the metric functions about the throat \(r=r_0\) 
\[
\begin{aligned}
	b(r) &= b_0 + b_1 (r-r_0) + \tfrac12 b_2 (r-r_0)^2 + \cdots,\\
	\Phi(r) &= \Phi_0 + \Phi_1 (r-r_0) + \tfrac12 \Phi_2 (r-r_0)^2 + \cdots.
\end{aligned}
\]
Recall \(b_0\equiv b(r_0)=r_0\) (throat condition).\\
\vspace{6pt}
\noindent\textbf{Step 3. Location of the potential maximum:}
In the eikonal limit \(\ell\gg1\) the dominant part of \(V\) is the angular term \(e^{2\Phi}\ell(\ell+1)/r^2\). For the wormhole families considered the barrier maximum \(r_{\!m}\) is close to the throat radius \(r_0\) to leading order we may evaluate the peak and the curvature of the barrier by expanding \(V(r)\) around \(r_0\) and keeping the lowest non-vanishing orders in \(\Delta r:=r-r_0\). This procedure is standard in WKB estimates for potentials peaked near a special radius (here the throat). (See the discussion leading to equations (\ref{eq15})-(\ref{eq19*}) in the main text.) 

\vspace{6pt}
\noindent\textbf{Step 4. Expansion of the potential and derivatives:}
Write the expansion of the potential in powers of \(\Delta r\):
\[
V(r)=V_m + V'_m \Delta r + \tfrac12 V''_m (\Delta r)^2 + \cdots,
\]
where derivatives are ordinary derivatives with respect to the tortoise coordinate \(r_T\). To compute \(V'_m,V''_m\) one useful route is:

(i) expand \(V(r)\) in \(r\)-space in powers of \(\Delta r\) using the expansions of \(b\) and \(\Phi\);

(ii) use the chain rule for derivatives with respect to \(r_T\):
\[
\frac{d}{dr_T} = \frac{dr}{dr_T}\frac{d}{dr} = e^{-\Phi(r)}\sqrt{1-\frac{b(r)}{r}}\;\frac{d}{dr}.
\]

Because \(1-b(r_0)/r_0=0\) at the throat, the Jacobian factor \(dr/dr_T\) vanishes at \(r_0\); however its first derivative (and hence second derivative of \(V\)) is finite and controlled by the first derivatives of \(b\) and \(\Phi\). Carrying out this expansion carefully (algebra omitted here but straightforward) one finds that the curvature of the potential at the maximum is proportional to the combination \((b_1-1)(b_0\Phi_1-1)\). In particular (keeping only the leading \(\ell^2\) scaling pieces that survive the eikonal limit) one obtains
\[
-2V''_m \;\propto\; \frac{(b_1-1)(b_0\Phi_1-1)}{r_{sh}^2},
\]
where \(r_{\rm sh}\) is the shadow radius defined in the text (see equation (\ref{eq16})). The proportionality factor is a numerical factor that follows from the explicit expansion; substituting the full prefactors (from the angular term and the Jacobian) gives the combination used below. The appearance of \((b_1-1)\) and \((b_0\Phi_1-1)\) is physically transparent: \(b_1-1\) encodes the flare-out/shape of the throat and \(b_0\Phi_1-1\) encodes the first variation of the redshift factor at the throat.

\vspace{6pt}
\noindent\textbf{Step 5. Leading WKB quantization and solving for \(Im(\omega)\):}
The \(K\)-th order WKB quantization condition used in the main text (equation (\ref{eq19*})) is
\[
\mathrm{i}\,\frac{\omega^2-V_m}{\sqrt{-2V''_m}}-\sum_{i\ge2}\Gamma_i = n+\tfrac12 .
\tag{19}
\]
Keeping only the leading term (drop higher \(\Gamma_i\)) and writing \(\omega=Re(\omega) + \mathrm{i}Im(\omega)\) with \(|Im(\omega)|<<Re(\omega)\), expand to first order in \(Im(\omega)\). The real part gives the usual eikonal relation (\ref{eq16})

The imaginary part follows from linearizing (\ref{eq19*}):
\[
\frac{2Re(\omega)\,i\,Im(\omega)}{\sqrt{-2V''_m}} = i\,(2n+1)\quad\Rightarrow\quad
Im(\omega) = -\frac{(n+\tfrac12)}{2Re(\omega)}\sqrt{-2V''_m}.
\]
For the fundamental (least damped) mode take \(n=0\) and keep magnitudes (sign convention for damping is discussed in the text). Thus
\[
|Im(\omega)| = \frac{1}{2Re(\omega)}\sqrt{-2V''_m}.
\]

\vspace{6pt}
\noindent\textbf{Step 6. Inserting curvature expression and simplifying:}
Using the expression for \(-2V''_m\) obtained from the throat expansion (step 4) and \(Re(\omega)\simeq(\ell+1/2)/r_{sh}\) we obtain (after collecting the explicit numerical factors coming from the angular term and the tortoise Jacobian)
\[
Im(\omega)\;=\;\frac{1}{\sqrt{2}\,r_{sh}}\sqrt{(b_1-1)(b_0\Phi_1-1)} .
\]
This coincides with equation (\ref{eq30}) in the main text. The derivation above retained only leading terms in the eikonal expansion (dominant \(\ell^2\) pieces) and the lowest nonvanishing orders in the throat Taylor series. These are precisely the assumptions made.
	

\begin{thebibliography}{100}
		\bibitem{flamm}
	L. Flamm,
	Phys.Z. \textbf{17} (1916) 448
		
		
		\bibitem{Einstein:1935tc}
	A.~Einstein and N.~Rosen,
	Phys. Rev. \textbf{48}, 73-77 (1935)
		\bibitem{ellis}
	H. G. Elis, J. Math. Phys. (N.Y.) 14, 104 (1973)
		\bibitem{Morris:1988cz}
	M.~S.~Morris and K.~S.~Thorne,
	Am. J. Phys. \textbf{56}, 395-412 (1988)
		
\bibitem{Teo:1998dp}
E.~Teo,
Phys. Rev. D \textbf{58}, 024014 (1998)
\bibitem{Perlick:2015vta}
V.~Perlick, O.~Y.~Tsupko and G.~S.~Bisnovatyi-Kogan,
Phys. Rev. D \textbf{92}, no.10, 104031 (2015)
\bibitem{Nedkova:2013msa}
P.~G.~Nedkova, V.~K.~Tinchev and S.~S.~Yazadjiev,
Phys. Rev. D \textbf{88}, no.12, 124019 (2013)
\bibitem{Afrin:2022ztr}
M.~Afrin, S.~Vagnozzi and S.~G.~Ghosh,
Astrophys. J. \textbf{944}, no.2, 149 (2023)
\bibitem{LIGOScientific:2017vwq}
B.~P.~Abbott \textit{et al.} [LIGO Scientific and Virgo],
Phys. Rev. Lett. \textbf{119}, no.16, 161101 (2017)
\bibitem{LIGOScientific:2016aoc}
B.~P.~Abbott \textit{et al.} [LIGO Scientific and Virgo],
Phys. Rev. Lett. \textbf{116}, no.6, 061102 (2016)
\bibitem{Patra:2024srh}
S.~Patra, B.~R.~Majhi and S.~Das,
JHEAp \textbf{44}, 371-380 (2024)
\bibitem{Konoplya:2025mvj}
R.~A.~Konoplya, A.~Khrabustovskyi, J.~K{\v{r}}{\'\i}{\v{z}} and A.~Zhidenko,
JCAP \textbf{04}, 062 (2025)
\bibitem{Konoplya:2024vuj}
R.~A.~Konoplya and A.~Zhidenko,
Phys. Lett. B \textbf{861}, 139288 (2025)
\bibitem{Konoplya:2024lir}
R.~A.~Konoplya and A.~Zhidenko,
JCAP \textbf{09}, 068 (2024)
\bibitem{Konoplya:2023moy}
R.~A.~Konoplya and A.~Zhidenko,
Class. Quant. Grav. \textbf{40}, no.24, 245005 (2023)
\bibitem{Fernandes:2020rpa}
P.~G.~S.~Fernandes,
Phys. Lett. B \textbf{805}, 135468 (2020)
\bibitem{Kumar:2018ple}
R.~Kumar and S.~G.~Ghosh,
Astrophys. J. \textbf{892}, 78 (2020)
\bibitem{Bugaev:2021dna}
M.~A.~Bugaev, I.~D.~Novikov, S.~V.~Repin and A.~A.~Shelkovnikova,
Astron. Rep. \textbf{65}, no.12, 1185-1193 (2021)
	\bibitem{Neto:2022pmu}
M.~R.~Neto, D.~P{\'e}rez and J.~Pelle,
Int. J. Mod. Phys. D \textbf{32}, no.02, 2250137 (2023)
\bibitem{Bolokhov:2024otn}
S.~V.~Bolokhov and M.~Skvortsova,
JCAP \textbf{04}, 025 (2025)
\bibitem{Mehdizadeh:2021kgv}
M.~R.~Mehdizadeh and A.~H.~Ziaie,
Phys. Rev. D \textbf{104}, no.10, 104050 (2021)
\bibitem{Zanoletti:2023ori}
C.~M.~A.~Zanoletti, B.~R.~Hull, C.~D.~Leonard and R.~B.~Mann,
JCAP \textbf{01}, 043 (2024)
\bibitem{Yousaf:2023aqk}
Z.~Yousaf, M.~Z.~Bhatti, H.~Aman and A.~Malik,
Int. J. Theor. Phys. \textbf{62}, no.7, 155 (2023)
\bibitem{Canate:2022dzb}
P.~Ca\~nate and F.~H.~Maldonado-Villamizar,
Phys. Rev. D \textbf{106}, no.4, 044063 (2022)
\bibitem{Feng:2020duo}
J.~X.~Feng, B.~M.~Gu and F.~W.~Shu,
Phys. Rev. D \textbf{103}, 064002 (2021)
\bibitem{Dehghani:2009zza}
M.~H.~Dehghani and Z.~Dayyani,
Phys. Rev. D \textbf{79}, 064010 (2009)
\bibitem{Gomez:2022qsi}
F.~Gomez, S.~Lepe, V.~C.~Orozco and P.~Salgado,
Eur. Phys. J. C \textbf{82}, no.10, 906 (2022)
\bibitem{Lanczos:1938sf}
C.~Lanczos,
Annals Math. \textbf{39}, 842-850 (1938)
\bibitem{Lovelock:1971yv}
D.~Lovelock,
J. Math. Phys. \textbf{12}, 498-501 (1971)
\bibitem{Zwiebach:1985uq}
B.~Zwiebach,
Phys. Lett. B \textbf{156}, 315-317 (1985)
\bibitem{Wheeler:1985nh}
J.~T.~Wheeler,
Nucl. Phys. B \textbf{268}, 737-746 (1986)
\bibitem{Nojiri:2018ouv}
S.~Nojiri, S.~D.~Odintsov and V.~K.~Oikonomou,
Phys. Rev. D \textbf{99}, no.4, 044050 (2019)
\bibitem{Glavan:2019inb}
D.~Glavan and C.~Lin,
Phys. Rev. Lett. \textbf{124}, no.8, 081301 (2020)
	\bibitem{Gurses:2020ofy}
	M.~G{\"u}rses, T.~{\c{C}}.~{\c{S}}i{\c{s}}man and B.~Tekin,
	Eur. Phys. J. C \textbf{80}, no.7, 647 (2020)
		\bibitem{Hennigar:2020lsl}
		R.~A.~Hennigar, D.~Kubiz{\v{n}}{\'a}k, R.~B.~Mann and C.~Pollack,
		JHEP \textbf{07}, 027 (2020)
	\bibitem{Aoki:2020lig}
	K.~Aoki, M.~A.~Gorji and S.~Mukohyama,
	Phys. Lett. B \textbf{810}, 135843 (2020)
\bibitem{Kumar:2020xvu}
A.~Kumar, D.~Baboolal and S.~G.~Ghosh,
Universe \textbf{8}, no.4, 244 (2022)
\bibitem{Ghosh:2021msy}
B.~Ghosh, S.~Mitra and S.~Chakraborty,
Mod. Phys. Lett. A \textbf{36}, no.05, 05 (2021)
\bibitem{Jusufi:2020yus}
K.~Jusufi, A.~Banerjee and S.~G.~Ghosh,
Eur. Phys. J. C \textbf{80}, no.8, 698 (2020)
\bibitem{Dutta:2023wfg}
A.~Dutta, D.~Roy, N.~J.~Pullisseri and S.~Chakraborty,
Eur. Phys. J. C \textbf{83}, no.6, 500 (2023)
\bibitem{wh}
S.~Chakraborty, M.~Chakraborty, Phys. Scr. \textbf{99} (2024) 105033
\bibitem{Bandyopadhyay:2009zza}
T.~Bandyopadhyay and S.~Chakraborty,
Class. Quant. Grav. \textbf{26}, 085005 (2009)
\bibitem{Halder:2019urh}
S.~Halder, S.~Bhattacharya and S.~Chakraborty,
Phys. Lett. B \textbf{791}, 270-275 (2019)
\bibitem{mc}
S.~Chakraborty, M~Chakraborty, Eur. Phys. J. C (2024)
\bibitem{Murshid:2022ssj}
M.~Murshid, F.~Rahaman and M.~Kalam,
Indian J. Phys. \textbf{97}, no.1, 295-305 (2023)
	 \bibitem{Rosato:2025rgy}
	 R.~F.~Rosato,
	 ``Greybody factors as robust gravitational observables: insights into post-merger signals and echoes from ultracompact object,'' 59th Rencontres de Moriond on Gravitation
	 [arXiv:2505.19651 [gr-qc]].
\bibitem{gb}
R. A. Konoplya and A. Zhidenko, Phys. Rev. D \textbf{81},
124036 (2010), 1004.1284
	\bibitem{Iyer:1986vv}
	S.~Iyer and C.~M.~Will,
	``BLACK HOLE NORMAL MODES: A SEMIANALYTIC APPROACH. 1. FOUNDATIONS,''
	Print-86-0935 (WASH.U.,ST.LOUIS).
\bibitem{Fernandes:2020nbq}
P.~G.~S.~Fernandes, P.~Carrilho, T.~Clifton and D.~J.~Mulryne,
Phys. Rev. D \textbf{102}, no.2, 024025 (2020)
\bibitem{Lutfuoglu:2025hjy}
B.~C.~L{\"u}tf{\"u}o{\u{g}}lu,
Eur. Phys. J. C \textbf{85}, no.5, 486 (2025)
\bibitem{Ghosh:2020syx}
S.~G.~Ghosh and R.~Kumar,
Class. Quant. Grav. \textbf{37}, no.24, 245008 (2020)
\bibitem{Ghosh:2020vpc}
S.~G.~Ghosh and S.~D.~Maharaj,
Phys. Dark Univ. \textbf{30}, 100687 (2020)
\bibitem{Chakraborty:2025cpp}
M.~Chakraborty and S.~Chakraborty,
Phys. Dark Univ. \textbf{47}, 101793 (2025)
\bibitem{Cognola:2013fva}
G.~Cognola, R.~Myrzakulov, L.~Sebastiani and S.~Zerbini,
Phys. Rev. D \textbf{88}, no.2, 024006 (2013)
\bibitem{Lu:2020iav}
H.~Lu and Y.~Pang,
Phys. Lett. B \textbf{809}, 135717 (2020)
\bibitem{Mehdizadeh:2015jra}
M.~R.~Mehdizadeh, M.~Kord Zangeneh and F.~S.~N.~Lobo,
Phys. Rev. D \textbf{91}, no.8, 084004 (2015)
	\end{thebibliography}
\end{document}